\documentclass[preprint]{rmaa}
\usepackage{paralist}

\usepackage{psfrag,color}

\newcommand{\shape}{{\em SHAPE }}

\title{Morpho-kinematic modeling of gaseous nebulae with {\it SHAPE}}

\author{
  Wolfgang Steffen,\altaffilmark{1} 
  and Jos\'e Alberto L\'opez,\altaffilmark{1}
}

\altaffiltext{1}{Instituto de Astronom\'\i{}a, UNAM, Ensenada, M\'exico.}

\shortauthor{Steffen, L\'opez}
\shorttitle{Morpho-kinematic modeling of gaseous nebulae with {\it SHAPE}}

\fulladdresses{
\item Instituto de Astronom\'{\i}a, Universidad Nacional Aut\'onoma de
M\'exico, Apartado Postal 877, 22860 Ensenada, B.C., M\'exico
(wsteffen@astrosen.unam.mx).
}

\listofauthors{W. Steffen, J.A. L\'opez, H. Riesgo}
\indexauthor{Steffen, W.}
\indexauthor{L\'opez, J.A.}
\indexauthor{Riesgo, H.}
\abstract{We present a powerful new tool to analyse and disentangle the 3-D geometry and
kinematic structure of gaseous nebulae. The method consists in combining
commercially available digital animation software to simulate the 3-D structure
and expansion pattern of the nebula with a dedicated, purpose built
rendering software that produces the final images and long slit spectra
which are compared to the real data. We show results
for the complex planetary nebulae NGC~6369 and Abell~30 based on long slit
spectra obtained at the San Pedro M\'artir observatory. }

\resumen{Presentamos una nueva herramienta poderosa para analizar y
desenredar la geometr\'{\i}a
3-D y la estructura cinem\'atica de nebulosas gaseosas. El m\'etodo consiste
en combinar software comercial para animaci\'on digital para simular
la estructura 3-D y el modo de expansi\'on de la nebulosa junto con un software
de representaci\'on gr\'afica de imagenes y perfiles de l\'{\i}nea dise\~nado
especialmente para este prop\'osito. Las im\'agenes resultantes se comparan
directamente con datos reales. Presentamos resultados
para las complejas nebulosas planetarias NGC~6369 y Abell~30 basados en
espectros de rendija larga obtenidos en el observatorio de San Pedro M\'artir.}

\addkeyword{ISM: modeling}
\addkeyword{Planetary nebulae: individual (NGC6369), individual (Abell 30)}
\addkeyword{Stars: Mass loss}

\begin{document}
% Typeset article header
\maketitle

\section{Introduction}
\label{sec:introduction}

In recent years, the discovery of a variety of complex structures in
planetary nebulae has opened many questions regarding the origin and evolution
of these objects (e.g. L\'opez 2000). Deviations from simple expanding shells
can include collimated outflows, poly-polar and
point-symmetric structures, rings or disks. These observations have lead to
a wealth of theoretical research into the effects of stellar magnetic fields,
rapid rotation and binarity of the central stars, and their evolutionary path
from spherically symmetric to bipolar mass-loss (e.g. Balick \& Frank, 2002,
and references therein).
In the absence of spherical symmetry, the tilt of the nebula with respect to
the line of sight and the location and position angle of the slit on the
nebula can often result in complicated position-velocity (P-V) diagrams
that can be difficult to interpret. The correct interpretation of
the nebular 3-D geometry and kinematic structure of PNe
is key to the understanding of the dynamics ruling their origin and evolution.

Modeling of line emission intensity maps have been used to obtain density
distributions over the face of the nebula in order to assess 3-D structures,
assuming pure photoionization from the central star (e.g. Monteiro et al. 2004;
Morisset, Stasi\'nska \& Pe\~na, 2005) but without incorporating kinematic
information or assuming simple velocity laws (e.g. Ragazzoni et al. 2001).

In this paper we present a new interactive 3-D modeling tool called \shape
which combines the versatility of commercial 3-D modeling software with a
rendering module specifically developed for application
in astrophysical research. The application of this method yields a 3-D
emissivity and velocity distribution for the object. Furthermore, different
velocity laws can be applied to different sections of the nebula to reproduce
the complex velocity patterns often observed in PNe. We exemplify the power
of this new method with models of the particularly complex planetary nebulae
Abell 30 and NGC 6369.

\section{Problem Definition}

The problem we attack with \shape is to characterize the current 3-D morphology
and velocity field of a nebula based on imagery and spectral kinematic
information. Detailed knowledge of this information leads to a better
understanding of the physical structure and dynamical evolution of a
gaseous nebula.

The projected image on the sky of an extended nebula provides
bidimensional spatial information of its structure. On the other hand,
the velocity field provides information on the radial component
of the velocity vector along the line of sight and conveys limited
but useful information on its depth or third spatial dimension.
However, an unambiguous solution of the complete 3-D structure at least
requires full knowledge of the velocity field. This situation is usually
not given, although topological and symmetry information
apparent in the images and spectra may help resolve ambiguities.
The simplest case occurs if the velocity of a volume element is constant over most of the
expansion time. In complex objects, this type of velocity distribution can
be expected if the nebula has evolved from a relatively short mass-loss
event and is now moving ballistically (e.g. Zijlstra et al. 2001) or from a continuous
interaction of a wind with small scale structures (Steffen \& L\'opez, 2004).
In these cases, after a sufficiently long time, the velocity pattern becomes
proportional to the distance from the center (a hubble-like velocity law).
The expansion of such a nebula is self-similar, i.e. the global shape is
conserved over time.

Under such conditions, the velocity vector is proportional to the position 
vector of every material element in the nebula. The shape of 
the nebula along the line of sight
is mapped linearly into the corresponding component of the velocity vector.
Hence, the Doppler-shift, which is equivalent to the velocity component
along the line of sight, is a map of the structure that is lost in a 
direct image, i.e. the long-slit spectrum allows a
view of the nebula from a direction perpendicular to the line of sight.
This situation can clearly be appreciated in the case of axisymmetric bipolar
nebulae, where the line profiles also show a bipolar structure that represents
the depth or third dimension of the 2-D image on the sky.
In many planetary nebulae an expansion velocity proportional to distance
from central star seems to be a reasonable
approximation at least for the brightest regions (e.g. Wilson 1950, Weedman 1968).
However, more complex velocity structures can
be expected when one or more mass loss events arise over a significant timescale
compared with the age of the nebula.

Sabbadin et al. (2000)
have used the assumption of a radial velocity field proportional to the
distance from the central star to reconstruct the 3-D structure of a
number of nebulae with a "tomographic" method. This tomography works
well as long as there are no significant deviations from the
hubble-like velocity law.

\section{The 3D modeling system}

In this section we
describe our new version of the code \shape as a tool to find the 3-D structure
and kinematics of gaseous nebulae.

Originally \shape was based on a description of structure and kinematics
using parametric geometrical equations on a regular 3-D grid
(Steffen et al. 1996). This code was adapted with a simple graphical interface
by Harman et al. (2004). It has been applied to a variety of objects
from individual knots in planetary nebulae (L\'opez, Steffen \& Meaburn, 1997)
to moderately complex structures in active galactic nuclei (Steffen et al. 1996).
In the present upgraded version we have devised a completely different
approach based on particle systems, rather than a regular grid, in combination
with a commercial 3-D animation package, which we describe next.

As our 3-D modeling software we use {\em Autodesk 3DStudio Max 7} ({\em
3DStudio Max} is a trademark of {\em Autodesk Media and Entertainment}, see
the website {\em www.discreet.com} for detailed software information).
We apply the available tools of this software to create
a particle and velocity distribution in space and time in order
to model an object. In particular we use the {\it ParticleFlow} particle
system to generate particle distributions which are then exported and
rendered in \shape.

\shape renders images and spectral information from the kinematics of the model
particle distribution. Key parameters such as orientation of the object on the 
sky, location and width of the slit, seeing values and spectral and spatial 
resolution are handled interactively in the graphical interface of \shape.

The general modeling process is as follows. 
With the inspection of available observations
one obtains an initial rough idea of the structure and topology of the
object, which is then reproduced in the modeling software. For this
purpose one produces a distribution of particles in space with its corresponding
velocities. The particles may be distributed over a topologically complex surface
or throughout a volume. The resulting emissivity is integrated along the line
of sight. The modeling software allows a very complex object to be
built and since all the features may be variable in time, the time
evolution of the object may also be explored.

As a guide during modeling, a spectal preview feature has been developed.
For a limited number of particles it allows a rough version of
the P-V diagrams to be visualized during modeling
in the viewport of {\em 3DStudio Max}.

The particle data are then exported into an external file.
This file is read by the core code of \shape, which produces the rendered image
and corresponding P-V diagram.
In contrast to the real-time feature described above, the core code of \shape is
not limited by the number of particles. It has been tested with up to one
million particles. The rendered images are then compared with the observations.
To improve a model, changes may be introduced at any point of the procedure
until, after a number of iterations, a satisfactory approximation to
the observations are obtained. Alternative solutions to an ambiguous dataset
will not be automatically found, but may be sought for separately.

The processing of the data, including the viewing direction, the emissivity,
internal structure, size and other parameters of individual particles is
controled via a dedicated graphical interface. This interface has been programmed
as a module of the modeling software {\em 3DStudio Max} and is fully integrated in
its interface.

\begin{figure}[t]
  \includegraphics[width=0.485\columnwidth]{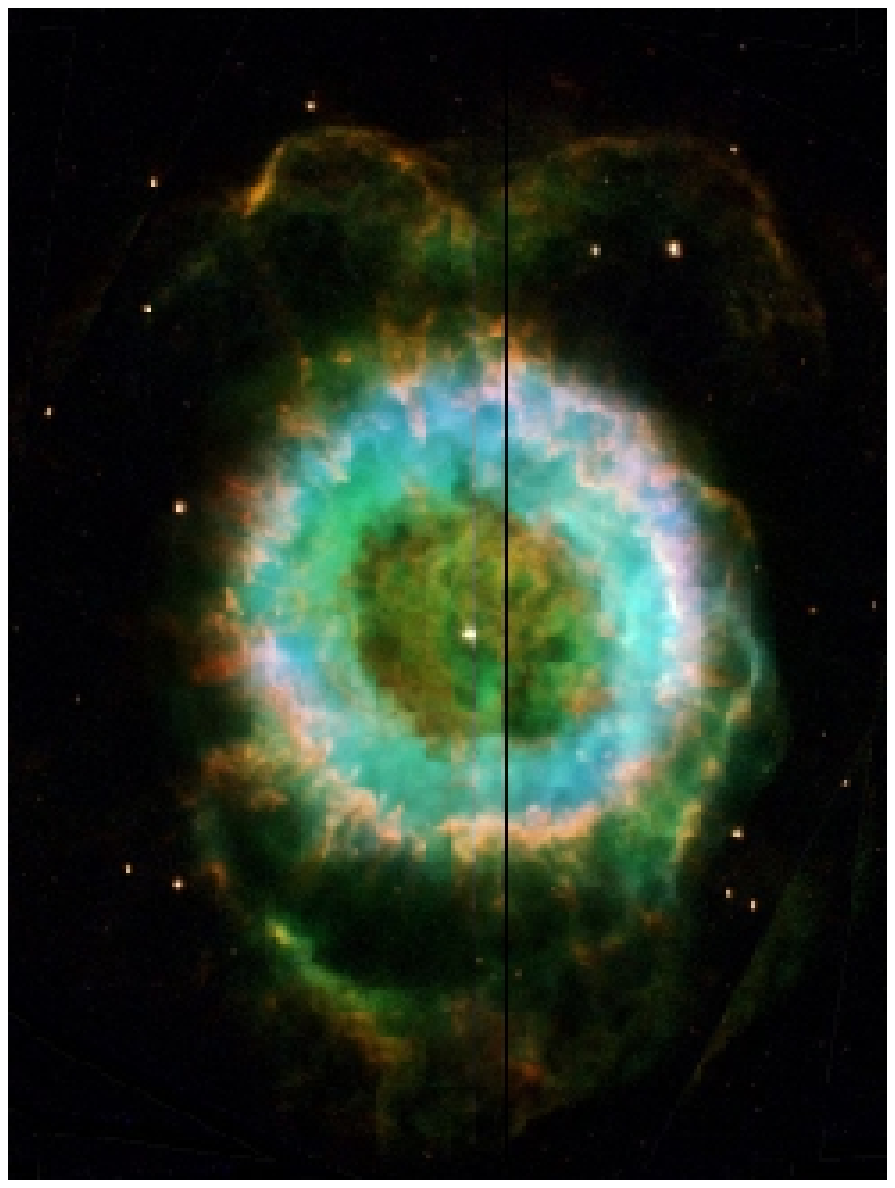}
  \includegraphics[width=0.49\columnwidth] {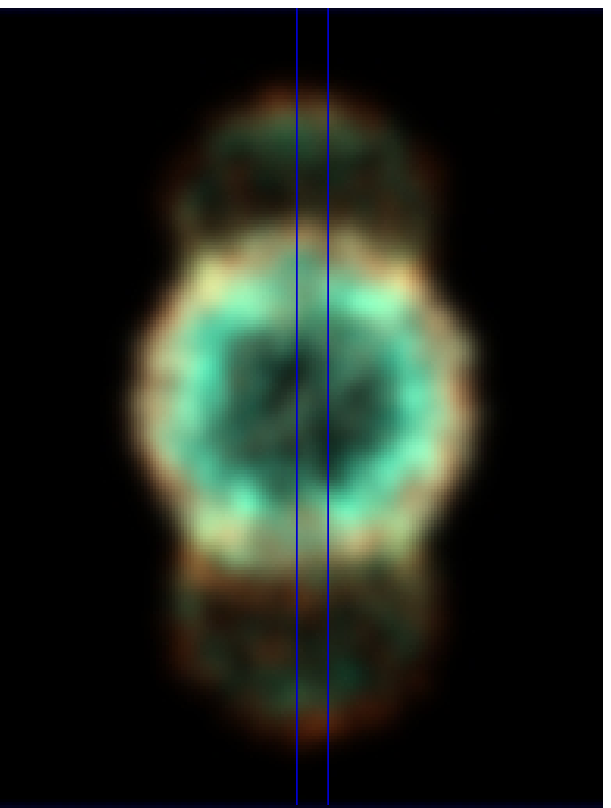}
  \caption{Hubble Space Telescope observation of NGC~6369 (left,
  Hubble Heritage Team, STScI, NASA) and
  our \shape image composing the H$\alpha$ in green/blue with NII in orange
  rendered at a resolution corresponding to approximately 0.3~arcsec.
  The long-slit position is marked with two vertical lines.
  }
  \label{fig:images}
\end{figure}

An important analytic feature of \shape is the possibility to apply different colors to
sections of a complex object. This allows a clear distinction
of them in the P-V diagrams, which helps considerably in the interpretation of
the observed spectra. By combining models of various emission lines, color images
are obtained to be compared with similar images from observations 
(see Figure \ref{fig:images}). A red-blue coloring mode allows a clear
distinction between red and blue-shifted regions in the model image.

Sequences of varying parameters like slit position and width, as well as
orientation of the object allow a systematic search for the
best parameters that match the observations or the production of image
sequences for animations which help considerably to visualize the 3-D structure of
the object. The P-V diagrams of objects that have been modeled as evolving in
time can also be visualized as a time sequence in animation form.

The most important current limitation of \shape is that only optically thin
nebulae can be modeled. The code does not perform any physical
radiation transport or line emission calculation.
What it does is to directly assign a relative emissivity
distribution, which is sufficient for its main purpose, the characterization
of the structure and kinematics of a nebula. Moreover, \shape can also be
used to produce complex density distributions and kinematics from which
photoionization models can be calculated using codes like NEBU\_3D (Morisset et
al, 2005). The current apparent drawbacks can hence be largely overcome by
combining \shape with codes which calculate radiation transport. We aim at
this next step in the near future.

\subsection{The rendering code}

The core code of \shape renders images and P-V diagrams from the position and
velocity data provided with {\em 3DStudio Max}. Each particle is read from a file
and its emission mapped to the image and spectrum. The intensity and color of
the emission from a particle may depend on the position in the object or on the
substructure of which it is part. The particles may also have a finite size larger
than single image pixels, as a well as an internal emissivity structure.
At this time this internal emissivity distribution may be either constant,
an exponential or a gaussian fall-off. This is useful mostly
for the initial modeling stages, before the convolution with the seeing parameters
are applied, because convolution eliminates all internal particle structure.
Particle sizes smaller than the seeing together with a sufficiently large number
of particles ensure adequate sampling of the object structure.

\begin{figure}[!t]\centering
  \includegraphics[width=\columnwidth]{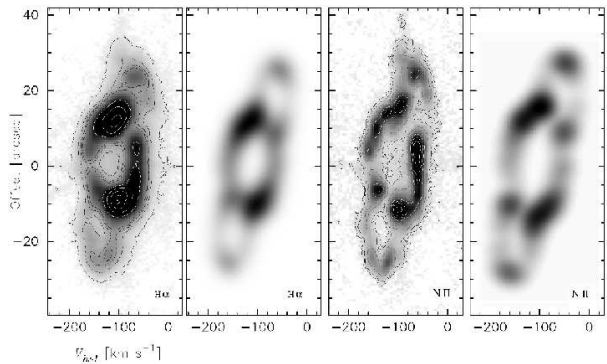}
  \caption{The observed long-slit emission line profiles of NGC~6369
(see Fig.\ref{fig:images}) are shown with contours overlayed for H$\alpha$ and NII 
together with the
  corresponding long-slit spectra resulting from the \shape modeling on the right of
each observed P-V diagram. Square-root gray-scales have been used for all images.}
  \label{fig:spectra}
\end{figure}

The image blurring due to seeing is modeled by convolution with a gaussian point-spread
function and including emission from within one
FWHM of the seeing disk outside the spectrograph slit. The contribution of this
emission decreases with distance from the edge of the slit.
The instrumental resolution is included
in the convolution of the raw image and P-V diagram with gaussian kernals of FWHM
corresponding to the spatial and velocity resolution of the instruments.
A Fast-Fourier-Transform algorithm is applied to calculate the convolutions
with gaussian kernels. After convolution, the P-V diagrams can be used to obtain
one-dimensional line profiles adding all or sections of the emission along the 
spacial domain, which are often useful
in work with low spatial resolution on the nebulae.
The core code of \shape has been programmed updating the Fortran code of the
original version.

In the following sections we present two example
models to reproduce observations of objects with different degrees of
complexity. An additional good example for an application of \shape
on the complex structure of NGC~6302 can be found in Meaburn et al. (2005).

\section{Observations and models}

\subsection{NGC~6369}

Our model of NGC~6369 is based on a long-slit spectrum obtained
with MES (Meaburn et al. 2003) on
the 2.1 m telescope at the San Pedro Martir Observatory. The spectral
resolution is 10 km s$^{-1}$ and the seeing 1.5 arcsec. The image has been
obtained from the Hubble Heritage Team, NASA, STScI (Figure \ref{fig:images}).
The spectrograph slit was located in the east-west direction, which
is coincident with the main axis of the
object (as indicated in the model image in Figure \ref{fig:images}).
The slit width corresponds to 1.5 arcsec on the sky.

\begin{figure}[t]\centering
  \includegraphics[width=0.9\columnwidth]{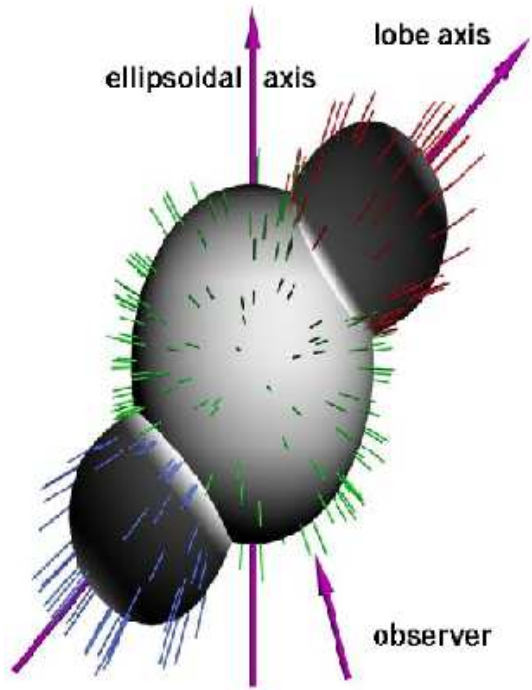}
  \caption{3-D structure of our model of NGC~6369 with velocity vectors shown
and basic emissivity mapped on the surface.}
  \label{fig:modvec}
\end{figure}

The observed spectra (Figure \ref{fig:spectra}, left) show that the central
"barrel" and the "lobes" are connected and probably conform a single topological
closed surface. Therefore we have started the modeling process with an initially
spherical shape and deformed it such that it matches our initial estimate
of the 3-D shape according to the image and spectrum. We then applied a brightness map
to the surface which approximates the emissivity distribution from
the H$\alpha$ and NII emission lines.
The structure and emission distribution is then adjusted interactively until
a satisfactory match was found. Figure \ref{fig:modvec} shows the
3-D structure of NGC~6369 depicted as a central ellipsoid with two opposite, 
off-axis, protruding lobes.

We have modeled the nebula's expansion with two different velocity laws.
Based on the observation that
the highest and lowest projected velocities in the spectra are similar for
the barrel and the lobes, we considered a constant expansion velocity directed
perpendicularly to the surface. This corresponds roughly to an energy driven
bubble (in this case three different bubbles, the barrel and the two lobes).
In this case we did not find any shape that would reasonably resemble
the image and spectra simultaneously.

The second model assumes that the velocity is proportional to distance and
the direction of motion is radial from the center of the nebula. The resulting
surface model with representative velocity vectors is shown in Figure
\ref{fig:modvec}. For the H$\alpha$ emission we assume that emission comes from
a region somewhat inside this surface and for NII slightly outside, with some
overlapping around the surface that is shown (as seen in the observed spectra and
images). The rendered image and spectrum are shown in Figures (\ref{fig:images}) and
(\ref{fig:spectra}), respectively. The particle number density is proportional to the
brightness on the 3-D surface (see Fig.\ref{fig:modvec}), which is not
accurately represented in the image due to lighting effects of this
visualization, which emphasizes the topology of the 3-D structure.

Our modeling indicates, that the line of sight is close to
the ellipsoid's symmetry axis (tilt $\approx 15^\circ$)
and the axis of the lobes is tilted approximately $40^\circ\pm 10^\circ$
with respect to that axis. The ratio between the height and the
diameter of the barrel is not well constrained by the current modeling and
observations. We estimate that this ratio is of the order $3/2$. The model images
and P-V diagrams have spatial and spectral resolution corresponding to those
in our observations, which are 1.5 arcsec and 10 km s$^{-1}$ using a slitwidth of
1.5 arcsec in the simulation.

We find that the model with the velocity law proportional to distance
produces an acceptable fit to the observed image and spectrum.
This suggests that this flow is in a relaxed momentum driven state; though
the lobes appear to move slightly slower than expected in this case
(Figure \ref{fig:spectra}). This might be an indication for a short-lived
collimated outflow, which is not acting on the lobes anymore,
and/or a stronger interaction of the
lobes with the ambient medium due to their lower densities.

Monteiro et al. (2004) proposed a ``diabolo''-type structure for
NGC~6369 based on the analysis of spectral imaging data of a number of
ions. The diabolo model in this case necessarily implies a narrow waist
in the equatorial plane with its symmetry axis close to the line of sight.
With this orientation such a narrow torus-like waist
is expected to have very low velocity along the line of sight, and
in the H$\alpha$ long\-slit spectrum it should appear as a narrow feature
near the systemic velocity. The observations presented in Figure
\ref{fig:spectra} do not show evidence for a narrow low-velocity waist.
However, bright emission regions near the systemic velocity in the
line profiles are apparent. In our model this is due to both, an
intrinsic equatorial density enhancement as seen in the image (Figure \ref{fig:images})
as well as a long tangential line of sight through the spheroidal main nebula.
Thus, our model does not require a waist to reproduce this feature.
NGC~6886 and NGC~6565 are two remarkably similar nebulae to NGC~6369,
both in terms of morphology and line-profile structure. Turatto et al. (2002)
have modeled the structure of NGC~6565 with their tomographic method obtaining
similar results to ours.

\subsection{Abell 30}

\begin{figure}[t]
\includegraphics[width=0.50\columnwidth]{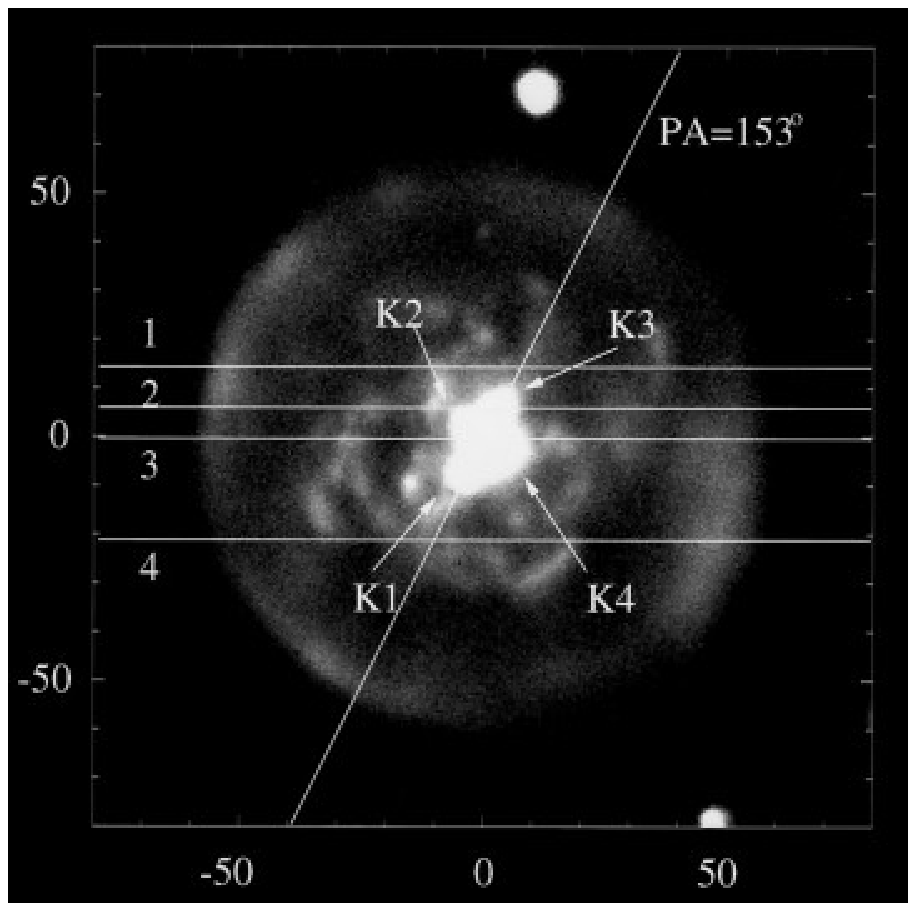} \includegraphics[width=0.47\columnwidth]{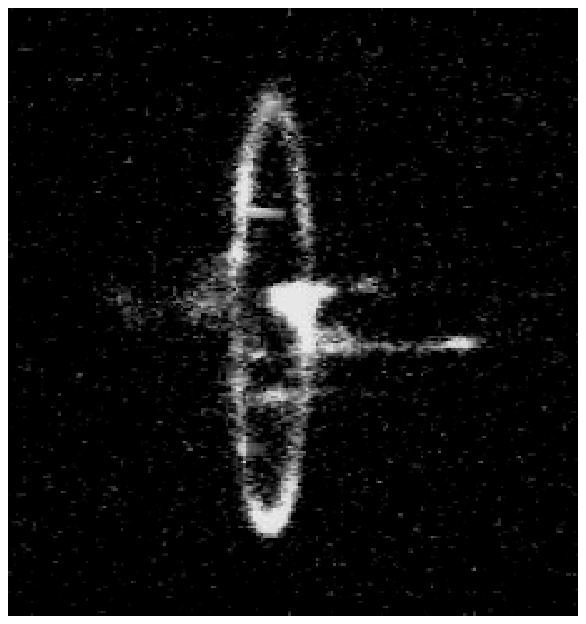}
   \caption{An image and P-V diagram of Abell~30 is shown for comparison with the model
  shown in Figure \ref{fig:abell30}. This figure is adapted from figures 1 and 2a of
  Meaburn \& L\'opez (1996, ApJ 472, L45) with permission of the authors.
  }
  \label{fig:abell30_obs}
\end{figure}

\begin{figure}[t]
 \includegraphics[width=0.48\columnwidth]{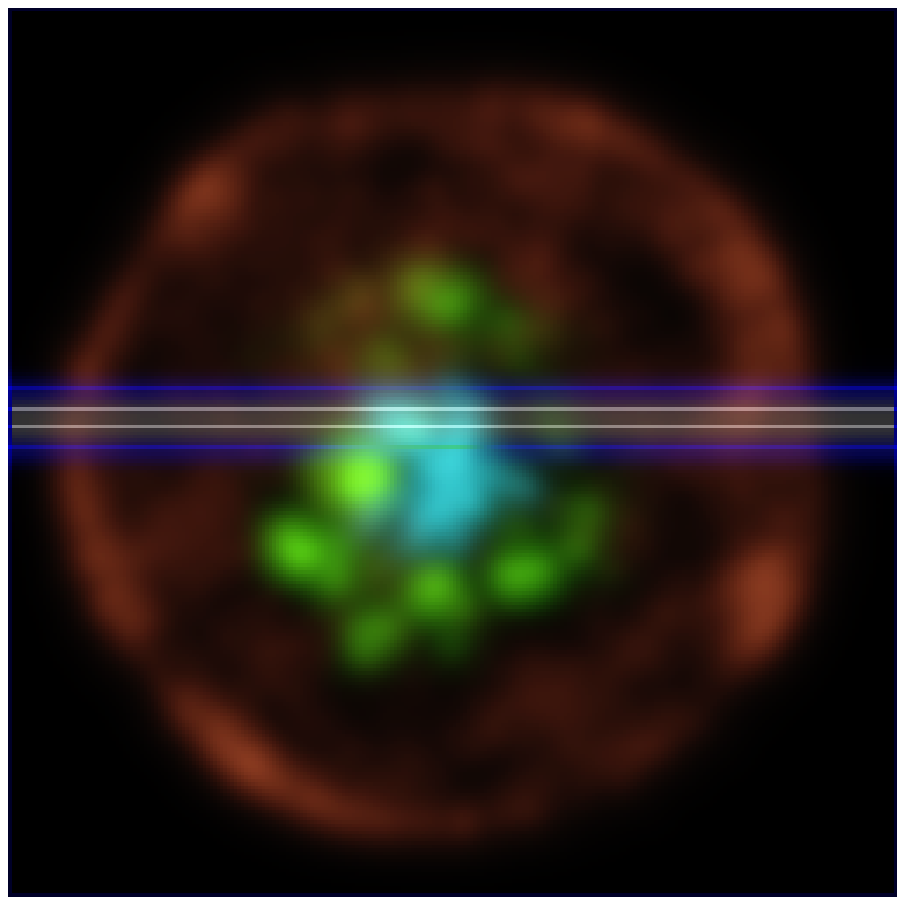}
 \includegraphics[width=0.48\columnwidth]{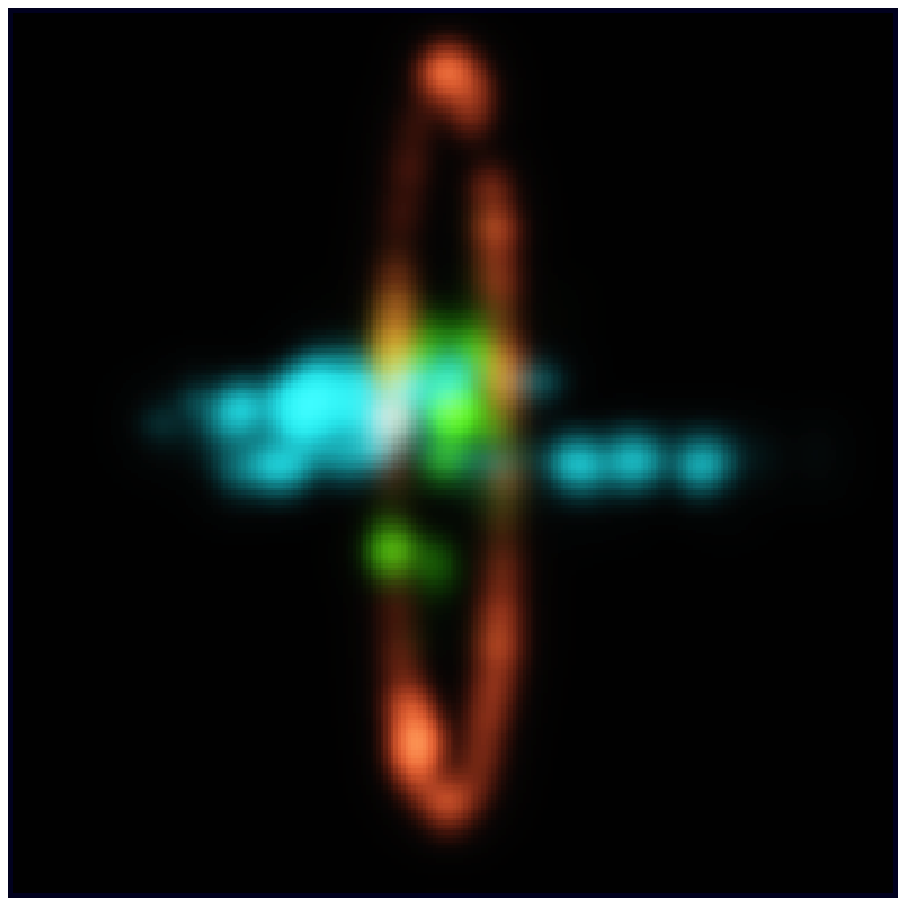}
  \caption{The image (left) and P-V diagram (right, position vertical and
  velocity horizontal) of a model for the
  planetary nebula Abell 30. The colors allow to identify different
  substructures in the image with those in the P-V diagram. The outer
  spherical shell is red, the intermediate knotty filamentas are green
  and the central cometary knots with their tails are blue. The vertical
  lines represent the coverage of the spectrograph slit with the range
  of seeing influence (outer slits).
  }
  \label{fig:abell30}
\end{figure}

Abell~30 is a hydrogen deficient planetary nebula with a spherical
[OIII] shell and complex knotty and filamentary structures in the
central region. The inner region contains
unusual cometary knots with large velocity spikes in the
P-V diagrams. Meaburn \& L\'opez (1996) describe its kinematics as "dramatic".
In this section we present a model produced with \shape as a
further illustration of its flexibility to handle complex kinematic
structures which may be very different from a hubble-like velocity law.
Observations of Abell~30 have been adopted from figures 1 and 2a in 
Meaburn \& L\'opez (1996) which are reproduced here in
Figure \ref{fig:abell30_obs}. The observations may be compared with our model
model image and P-V diagram for one slit position as shown in Figure
(\ref{fig:abell30}). 

The outer shell was modeled as a sphere with some variations in brightness and
a constant radial expansion. A random distribution of knots
and filaments was used for the inner regions.  For the cometary knots,
particles where emitted from discrete points which then interacted with a
central spherically symmetric wind, accelerating the particles outwards
and producing the high-speed features in the P-V diagrams. Support for the
existence of such a wind in Abell~30 comes from observations of X-rays
(Chu, Chang \& Conway, 1997).

For the model we adapted the parameters
similar to those of the observations with seeing between 1 and 2 arcsec and
a width of the slit of 1.9 arcsec. For the model we used 1.5 and 1.9 arcsec
respectively. Since the object is basically spherically symmetric, except
for the random distribution of brightness variations of the outer shell and
the distribution of knots in the central region, the specific orientation
of the model is not a fundamentally important parameter. 

At this time no attempt was
made to match individual small-scale features with those in Abell~30.
Still, the spectrum in Figure \ref{fig:abell30} matches the observed one
very well. This model shows how different methods can be used to combine 
substructures which have a variety of complexity and kinematic signatures.

\section{Discussion and Conclusions}

In this paper we have presented the upgraded version of \shape, with a
novel approach to the determination of the
three-dimensional structure and kinematics of gaseous nebula.
\shape combines the
capabilities of commercial 3-D modeling software with a purpose-built rendering
software and graphical control interface for its application to astrophysical
nebulae. \shape produces images and P-V diagrams for highly complex nebulae. The
results can be compared directly with observations and help to understand
structure, kinematics and orientation of the objects.

The kinematics as observed
from long-slit spectra often provides sufficient information about the
3-D structure and topology of an object for the modeling with \shape to
provide a self-consistent 3-D structure. In addition the \shape models 
can be applied directly as input density distributions for photo-ionization codes. 
In \shape one can define and combine different velocity laws that fit complex
structures. Solutions are expected to be most reliable if the object shows evidence
for a significant degree of symmetry, then the full 3-D structure and
kinematics can be deduced unambiguously.
In other cases the object may be devided into regions or subsystens
which allows them to be solved separately.

We are currently building a catalogue of of synthetic emission line profiles
with \shape that should be a useful reference to interpret long-slit observations
of PNe with diverse morphologies and orientations.

In this paper we showed applications of \shape to the planetary nebulae
Abell~30 and NGC~6369. As a result of our application of \shape to new
observations of the planetary nebula NGC~6369, we propose that its basic
structure is that of a ellipsoidal or barrel-shaped main nebula with bipolar
protrusions at a large angle to the symmetry-axis of the main nebula.
Abell~30 has been modeled combining different velocity patterns to fit the
expanding shell and the inner high-velocity knots.

\acknowledgements

We acknowledge support from DGAPA-UNAM projects IN111803 and IN112103 as well as
CONACYT projects 37214 and 43121.

\end{document}